\documentclass[12pt]{iopart}
\usepackage{bm}
\usepackage{hyperref}
\usepackage{framed}
\usepackage{subfigure}
\usepackage{hyperref}   
\usepackage{graphicx}
\usepackage{iopams}
\expandafter\let\csname equation*\endcsname\relax 
\expandafter\let\csname endequation*\endcsname\relax 
\usepackage{amsmath}
\usepackage{color}

\newcommand{\eq}[1]{\eref{#1}}

\renewcommand{\vec}[1]{\mathbf{#1}}
\newcommand{\matD}{\mathcal{D}_t}
\newcommand{\p}{\partial}

\usepackage[normalem]{ulem}

\begin{document}

\title{Analogue Black Hole Spectroscopy; or, how to listen to dumb holes}

\author{Theo Torres}
\address{School of Mathematical Sciences, University of Nottingham, University Park, Nottingham, NG7
2RD, UK}

\author{Sam Patrick}
\address{School of Mathematical Sciences, University of Nottingham, University Park, Nottingham, NG7
2RD, UK}

\author{Maur\'icio Richartz}
\address{Centro de Matem\'atica, Computa\c c\~ao e Cogni\c c\~ao,
Universidade Federal do ABC (UFABC), 09210-170 Santo Andr\'e, S\~ao Paulo, Brazil}

\author{Silke Weinfurtner}
\address{School of Mathematical Sciences, University of Nottingham, University Park, Nottingham, NG7
2RD, UK}
\address{School of Physics and Astronomy, University of Nottingham, Nottingham, NG7 2RD, UK.}
\address{Centre for the Mathematics and Theoretical Physics of Quantum Non-Equilibrium Systems, University of Nottingham, Nottingham, NG7 2RD, UK.}

\date{\today}
	\begin{abstract}
		Spectroscopy is a fundamental tool in science which consists in studying the response of a system as a function of frequency. Among its many applications in Physics, Biology, Chemistry and other fields, the possibility of identifying objects and structures through their emission spectra is remarkable and incredibly useful. In this paper we apply the spectroscopy idea to a numerically simulated hydrodynamical flow, with the goal of developing a new, non-invasive flow measurement technique. Our focus lies on an irrotational draining vortex, which can be seen, under specific conditions, as the analogue of a rotating black hole (historically named a dumb hole). 
		This paper is a development of a recent experiment that suggests that irrotational vortices and rotating black holes share a common relaxation process, known as the \emph{ringdown} phase.	
		 We apply techniques borrowed from black hole physics to identify vortex flows from their characteristic spectrum emitted during this ringdown phase. 
		 We believe that this technique is a new facet of the fluid-gravity analogy and constitutes a promising way to investigate experimentally vortex flows in fluids and superfluids alike.
		  
	\end{abstract}
	\maketitle

\section{Introduction}

One of the most surprising discoveries of modern physics is that, once quantum effects are taken into account, black holes are not completely dark. 
Rather they emit particles as hot bodies do, with specific temperatures linked to their masses~\cite{Hawking74}. 
Another important feature, which is true even at the classical level, is that they are also not quiet. 
Black holes in the sky, like guitar strings, ring when they are plucked. 
 While the sound made by the guitar is determined by the properties of the guitar (i.e.~the length and the density of the string plucked), the \emph{sound}~\footnote{Note that we use the term sound as a generic term to describe the frequency of generic radiation such as scalar, electromagnetic or gravitational waves.}  of a black hole depends on the characteristics (mass and angular momentum) of the black hole~\cite{Kokkotas99,Berti09}.
Additionally, in the same way that the guitar string does not vibrate forever due to dissipation, a black hole in an infinite space will eventually stop ringing.
This damped ringing stage after the black hole is plucked (or perturbed) is known as the ringdown phase. 
Since this ringdown phase is independent of the initial source of the perturbation, people have toyed with the possibility of identifying astrophysical black holes by listening to them~\cite{Press72, Echeverria}. 
While we still haven't seen black holes shine, recently the LIGO collaboration has heard them sing~\cite{LIGO_GW}. 
Therefore, one of the many achievements of LIGO has been to contribute to turn this perfect astrophysical pitch idea into a reality~\cite{LIGO_propBH}.

In 1981, less than a decade after Hawking discovered that black holes do radiate, Unruh suggested the use of sound waves in a flowing fluid to mimick black hole radiation in the laboratory. Unruh's conclusion was reached after he found out that the equation of motion governing the propagation of sound in a moving fluid and the wave equation for a massless scalar field propagating on a curved space-time are the same provided that certain assumptions are made~\cite{Unruh80}. Unruh had thus shown that the propagation of sound waves in water can be associated with the metric of a curved spacetime which is fully characterized by the background flow on top of which the waves propagate. In particular, by tuning the velocity of a fluid, one can simulate various curved space-times in the laboratory. For instance, when the fluid is flowing slower than the speed of sound in one region and faster in another, sound waves will experience the analogue of a black hole horizon. 
The corresponding region of the flow from which sound waves cannot escape has been named a dumb hole.

Unruh's fluid-gravity analogy, which had been originally designed to observe the radiation from a dumb hole horizon, was later extended to reproduce other gravitational phenomena in many different physical systems~\cite{Barcelo05}. More specifically,
in addition to Hawking radiation~\cite{Silke10,Euve16,Steinhauer16}, successful implementations of analogue gravity experiments have shed light onto fundamental phenomena such as superradiance~\cite{Torres16} and cosmological particle production~\cite{Fedichev03,Wittemer19}. 
Concurrently, these experimental studies have been a source of inspiration to develop theoretical tools to better understand these universal processes (see for example~\cite{Corley96,Unruh05,Richartz12,Richartz13,Coutant16} or~\cite{Barcelo05} and references therein).
Recently, a new analogue gravity experiment based on surface water waves demonstrated that the relaxation process of vortices can be understood using theoretical tools originally developed to describe the ringdown phase of perturbed black holes~\cite{Torres18}.
By looking at the characteristic waves emitted while the vortex is relaxing towards equilibrium, its authors were able to infer details about the fluid flow itself.

This idea of identifying objects via their characteristic response to perturbations is not only applicable to black holes or vortices. It is at the heart of the idea of spectroscopy. 
It also has a long mathematical history which started with the famous question: \emph{``Can one hear the shape of a drum?"}~\cite{Kac66} (in other words, is it possible to deduce the shape of a drum only from the information carried by the sound it emits?).   
Whilst the answer was quickly found to be ``no" in 16 dimensions, the negative answer for the more common 2-dimensional drum came 26 years later when two different planar regions emitting the same sound were constructed mathematically. However, if one restricts the shape of the drum then the answer to the question is ``yes". 

Inspired by the concept of spectroscopy, we pursue here the idea suggested in~\cite{Torres18}and develop a scheme to characterise irrotational vortex flows directly from their characteristic \emph{sound}.
We believe that the proposed scheme is a novel and promising way to measure fluid flows in both fluids and superfluids. We note that our method is different from previous non-invasive flow measurement techniques via wave scattering (see for e.g.~\cite{Steinberg97,Coutant15}), as we suggest to use the characteristic modes of the vortex itself instead of the properties of a scattered wave (such as the wavefront dislocation in the mentioned studies).

The paper is organised as follows. In sections \ref{Hydrons} and \ref{Resonances}, we review the formalism and the existing results in the literature in order to fix the notation and make the paper self-consistent.
In section \ref{ABHS} we detail the Analogue Black Hole Spectroscopy (ABHS) concept and illustrate its applicability in the case of an irrotational flow in the deep water regime.
In section \ref{Application} we present a detailed application of the ABHS method to extract flow parameters in the shallow water regime. This is followed by a discussion, section \ref{sec:disc}, which concludes the paper.
\section{Surface waves and Hydron Trajectory}\label{Hydrons}

Consider surface water waves propagating on an irrotational, incompressible and inviscid fluid of constant depth $h$. 
The fluid is described by a velocity field $\vec{v}$ of which the waves are small perturbations, i.e. $\vec{v} = \vec{v}_0 + \delta\vec{v}$. 
Since the flow is irrotational, we can write $\delta \vec{v} = \vec{\nabla} \phi$, where $\phi$ is a scalar function that fully characterizes the velocity pertubations. 
The wave equation associated with $\phi$ is given by~\cite{Milewski96}:
\begin{equation}\label{wave_equation}
\matD^2 \phi - i\left( g\vec{\nabla} - \gamma \vec{\nabla}^3\right)\tanh(-ih\vec{\nabla}) \phi = 0,
\end{equation}
where $\matD = \p_t + \vec{v}_0.\vec{\nabla}$ is the material derivative, $g$ is the gravitational acceleration and $\gamma$ is the ratio between the surface tension and the density of the fluid. 
To keep the discussion general and simplify notation, we write \mbox{$F(-i\vec{\nabla}) = -i\left( g\vec{\nabla} + \gamma \vec{\nabla}^3 \right)\tanh(-ih\vec{\nabla})$}.
In the shallow water regime and when surface tension is neglected, equation \eq{wave_equation} can be rewritten as the Klein-Gordon equation for a massless scalar field propagating on a curved spacetime~\cite{SCH02} whose metric is fully determined by the background flow parameters~\cite{Unruh80,Barcelo05}.
\newline

Due to the infinite number of spatial derivatives involved (in the $\tanh(-ih\nabla)$ operator) and due to the fact that the background is not homogeneous, the wave equation~\eq{wave_equation} cannot be solved neither analytically nor numerically. Nevertheless we can look for its solutions with a gradient expansion method \cite{Buhler}.
 
We start by assuming that the velocity potential has the following form:
\begin{equation}
\phi = A(\vec{x}) e^{i S(\vec{x})/\epsilon},
\end{equation}
where $\epsilon$ is a small parameter accounting for the fact that the local phase $S$ is rapidly oscillating compared to the local amplitude $A$. We also assume that the background flow changes on scales much greater than the wavelength. If we further expand $A$ and $S$ in powers of $\epsilon$,then each term in the sums $A = \sum A_n \epsilon^n$ and $S = \sum S_n \epsilon^n$ can be found iteratively using \eq{wave_equation}.

The first term of the local phase expansion, $S_0$, satisfies: 
\begin{equation} \label{HJ_eq}
\left( \matD S_0 \right)^2 - F(\vec{\nabla}S_0) = 0.
\end{equation}
At this level, the wave can be seen as a coherent system of rays. 
Rays are parametrized curves $\vec{x}(\tau)$ associated with a load $\vec{k}(\vec{x})$ defined by an Hamiltonian  $H$ via $H(\vec{x},\vec{k}) =0$ and a variational equation: $\delta \int \vec{k} d\vec{x} =0$ with fixed end points (see \cite{Synge63} for details of the Hamiltonian method applied to water waves). 
At leading order, our solution is a family of rays satisfying the circulation condition:
\begin{equation} \label{circulation}
\int_C \vec{k} d\vec{x} = 0,
\end{equation}
where $C$ is any closed circuit.
This condition implies that $\vec{k}$ is the gradient of a \emph{phase} and therefore identifies the Hamiltonian equation $H(\vec{x},\vec{k}) = 0$ with the Hamilton-Jacobi equation \eq{HJ_eq}. 
Until now, $\vec{x}$ denoted similarly time and spatial coordinates. We now make the separation clear by defining $\omega = - \p_t S_0$ and letting $\vec{k}=\nabla S_0$.
The Hamiltonian for our system is therefore \cite{Torres2017}:

\begin{equation}
H(\vec{x},\vec{k},t,\omega) = - \frac{1}{2} \left(\omega - \vec{v}_0.\vec{k} \right)^2 + \frac{1}{2} F(k)
\end{equation}

The rays obtained from Hamilton's equations can be viewed as the world lines of some fictitious particles called \emph{hydrons}~\cite{Purser62}. 
This is in analogy with the null geodesics of curved space-time being the trajectories of massless particle. 
In the same way that geodesics are linked to many phenomena in curved space-time, hydron trajectories can be used to estimate various effects in fluid flows \cite{Berry_tsunami,Torres2017}. 

\section{Relaxation and resonances: from Black Holes to vortex flows} \label{Resonances}

It is known that the late time behaviour of waves propagating around black holes is described by a set of modes with complex frequencies called quasinormal modes (QNMs)~\cite{Kokkotas99,Berti09}. 
These modes oscillate with a frequency $\omega_{mn}$ and decay exponentially in time with a timescale $1/\Gamma_{mn}$ during what is called the \emph{ringdown phase}. 
The subscript $m$ refers to the azimuthal component of the perturbation and $n$ classifies the QNMs according to their decay time. 
An important feature of the ringdown phase is that it is entirely characterized by the black hole parameters, i.e.~mass, charge and angular momentum. 
Based on the fluid-gravity analogy, it has been predicted that irrotational vortex flows, sometimes also called Draining Bathtub flows (DBT flows), should also emit QNMs when perturbed~\cite{Berti04}.

DBT flows are characterised by a velocity field
\begin{equation} \label{DBT}
\vec{v}(r,\theta) = \frac{C}{r} \vec{e}_\theta - \frac{D}{r} \vec{e}_r,
\end{equation}
where $(r,\theta)$ are polar coordinates on the 2-dimensional unperturbed free surface, $C$ is the circulation parameter and $D$ is the drain parameter.
Shallow gravity waves propagating at speed $c=\sqrt{gh}$ on a DBT flow experience the analogue of a rotating black hole metric~\cite{SCH02,Dolan13}. 
The analogue horizon is located at $r_{\rm hor} = D/c$ and the analogue ergosphere is at $r_{\rm ergo} = \sqrt{C^2 + D^2}/c$. 
In this regime, the quasinormal mode spectrum of the DBT flow has been calculated using various techniques borrowed from black hole physics \cite{Berti04,Dolan12}. 
One such technique is based on the geodesic structure of the effective spacetime and the existence of \emph{light-rings} (LRs)~\cite{Cardoso09}. 

In a gravitational context, LRs are orbits of massless particles, such as photons. 
In the fluid context, we have seen that hydrons play the role of a fictitious massless particles. 
One can therefore extend the concept of LRs to the hydrodynamical system and define them as the orbits (i.e.~equilibrium points in the radial direction) of hydrons~\footnote{One could therefore rename light-ring as hydron-rings when talking about the hydrodynamical system. However we will keep the name light-ring to stay in close connection with black hole physics even though we are dealing with surface gravity waves.}.
As it is possible to define hydron trajectories in general, and not only in the shallow water regime, the QNM spectrum has been evaluated when dispersive effects cannot be neglected. This was accomplished with the help of the LR method in~\cite{Torres2017} by solving Hamilton's equations and imposing the equilibrium point conditions, namely
\begin{equation}\label{LR_cond}
\dot{r} = \frac{\p H}{\p k_r} = 0,
\quad
\dot{k_r} = -\frac{\p H}{\p r} = 0
\quad \mathrm{and} \quad
H=0.
\end{equation}

For every fixed $m$, solving \eq{LR_cond} gives a unique triplet $(r_\star,k_\star,\omega_\star)$ where $r_\star$ is the radius of the LR and $\omega_\star$ gives an estimate of the real part of the quasinormal mode frequency. The imaginary part of the spectrum can be estimated by computing the \emph{Lyapunov exponent} $\Lambda$ which measures the convergence/divergence of the rays near the orbit via:
\begin{equation} \label{eq:hess}
\Lambda = \frac{\sqrt{- \det[d^2H]}}{|\p_\omega H|},
\end{equation}
where $[d^2H]$ is the Hessian matrix of the Hamiltonian function. Since the Hamiltonian depends only on the background flow and on the fluid parameters, the solutions of equations \eqref{LR_cond}, \eqref{eq:hess} that determine the LRs modes spectrum are fully characterized by the flow parameters $C$, $D$ and $h$. In both the shallow and the deep water regimes, the solutions can be obtained analitically. For example, in the shallow water regime, we have the linear relation between $\omega_\star$ and $m$,
\begin{equation}\label{shallow_LR}
\omega_\star(m) = c^2 \frac{\sqrt{C^2 + D^2}}{ B^2_\pm} |m|
\end{equation}
where $B_\pm = \left(  2(C^2 + D^2) \mp 2C\sqrt{C^2 + D^2}  \right)^{1/2}$ and the $\pm$ sign corresponds to the sign of $m$.
This linear scaling between $\omega_\star$ and $m$ changes in the deep water regime to the cubic relation
\begin{equation}
\omega_\star(m) = \frac{3}{8} \left( \frac{4 g^2}{ \sqrt{B^2_\pm - D^2}} \right)^{1/3} |m|^{1/3}.
\end{equation}  

While it has been shown that QNMs govern the relaxation process of a black hole, a recent experiment suggest that the same happens for vortex flows~\cite{Torres18}.
Indeed, the LR modes correspond to the lowest energy modes capable of travelling across the entire system; they represent the most efficient way to radiate extra energy away from vortex flows. 
One therefore expects these LR modes to be present around any physical vortex as one can never attain perfect equilibrium. 
Inspired by the black hole spectroscopy idea, we propose to turn this seemingly negative feature into a novel and promising flow measurement technique that we describe in the next section.

\section{Analogue Black Hole Spectroscopy}\label{ABHS}

The important feature that the QNM spectrum depends only on the black hole parameters opens up the possibility of \emph{black hole spectroscopy} (BHS).
By measuring the QNM spectrum of a black hole, one should be able to recover the parameters of the black hole that produced the spectrum~\cite{Press72,Echeverria89,Schutz09}.
With the recent detection of gravitational waves \cite{LIGO_GW,LIGO_propBH}, we seem to be close to turning this idea into a reliable astronomy technique. 
We suggest here the use of a similar approach to characterise vortex flows in fluids and superfluids. We describe in this section the general procedure of what we will call \emph{Analogue Black Hole Spectroscopy} (ABHS). 
The ABHS scheme is as follow:
\begin{itemize}
\item[1)] Measure the perturbations around the vortex flow one wishes to characterise.
\item[2)] Decompose the perturbation into its azimuthal components and extract the characteristic frequency for each $m$ mode.
\item[3)] Perform a non-linear regression analysis to find the best match between the measured experimental spectrum and the theoretical predictions obtained by solving \eq{LR_cond}.
\end{itemize}

The key features of this process that consists in analysing the LR spectrum associated with the vortex flow are detailed below:  \\

\textit{Step 1 - Measurement:} The first step of the process is the detection of the perturbations on top of the vortex flow. 
To perform the measurement, the system needs to be in a quasi-stationary regime. 
This means that the flow should not vary over the timescale of the measurement but the flow should not be in its equilibrium state in order to stimulate the emission of characteristic modes. 
After this stage, the perturbations are contained in a single variable $\phi(t,\vec{x})$. 
For example, $\phi$ can represent the surface elevation in the case of surface waves in classical fluids or the atom density in a BEC.
\\

\textit{Step 2 - Characteristic spectrum:} This step consists in extracting the information contained in each azimuthal component $\phi(t,m,r)$ which is obtained from the original signal $\phi(t,\vec{x})$ through an angular Fourier transform.
Note that in order to perform this step and to be able to distinguish between positive and negative $m$'s, $\phi$ needs to be a complex field. 
The analytic representation of the real valued measurement can be constructed via the Hilbert transform. 
From $\phi(t,m,r)$ one can construct the Power Spectral Density (PSD) $P_m(\omega,r) \propto |\tilde{\phi}|^2$ for each azimuthal number, where $\tilde{\phi}$ is the time Fourier transform of $\phi$. 
The PSD's are then used to identify the characteristic frequencies.
As we are considering a quasi-stationary system, frequencies are conserved. 
Therefore, any signal emitted at frequency $\omega_0$ will appear as a constant line in the PSD plots at $\omega = \omega_0$ in the $(\omega, r)$ plane over a radial range corresponding to the region of propagation of the signal. 
As the characteristic modes are the lowest energy modes capable of propagating across the system, they will appear as the lowest constant frequency line stretching over the entire radial region. 
By looking for this line in every $P_m(\omega,r)$, one can define a radius independent spectrum $\omega_{\rm exp}(m)$.
\\

\begin{figure*}[t]
\begin{center}
\includegraphics[scale = 0.8, trim= 1cm 0 1cm 0]{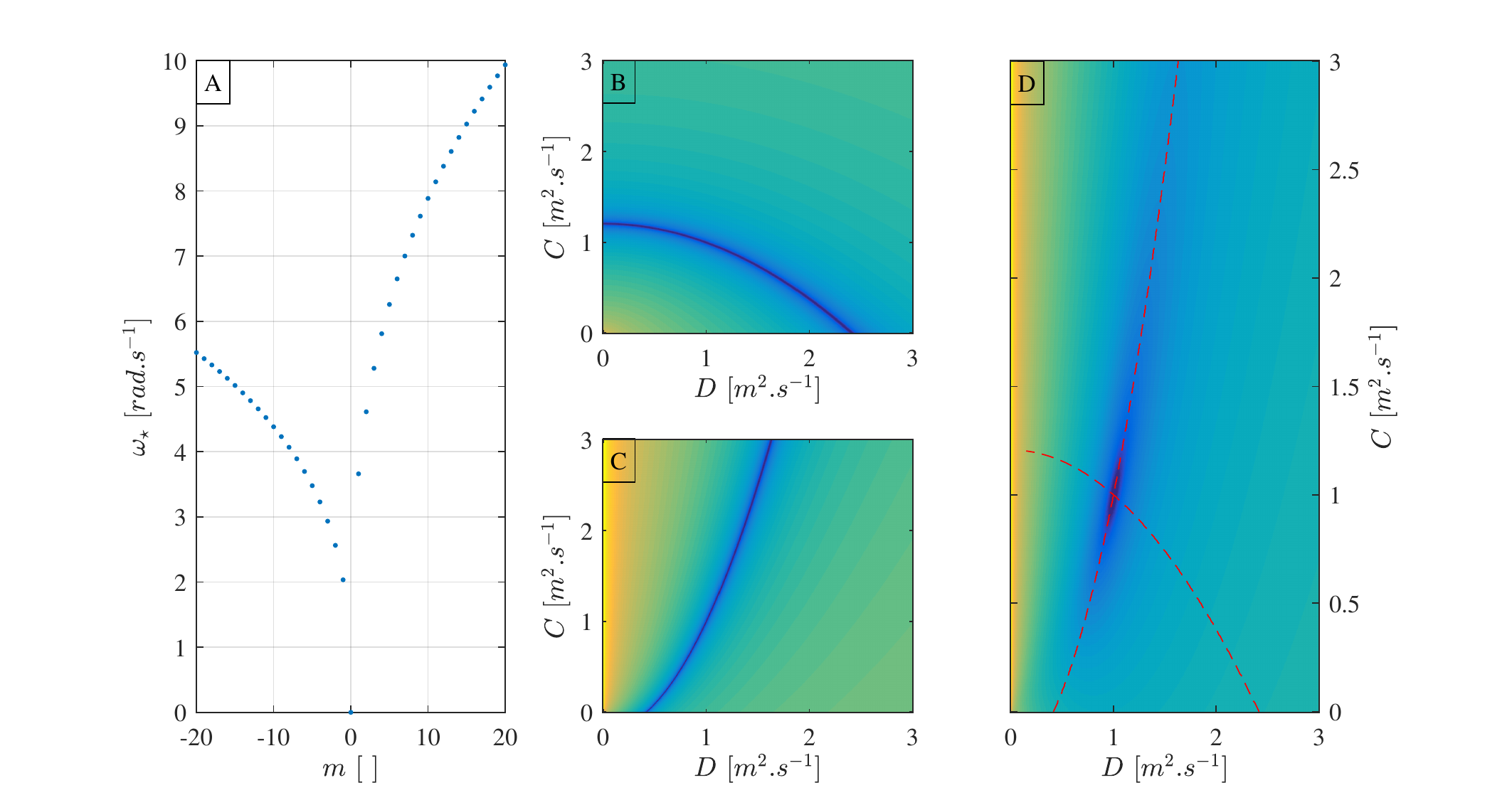}
\end{center}
\caption{\textbf{ Illustration of the Analogue Black Hole Spectroscopy in the deep water regime.}  
Panel A depicts the light-ring spectrum for surface waves in the deep water regime, i.e. $F(k) = gk$, computed with the flow parameters $C=D=1~m^2.s^{-1}$. Panel A (respectively B) represents the logarithm of the mean square error for a region of the $(C,D)$ parameter space computed using only the negative (respectively positive) $m$ part of the light-ring spectrum. Darker regions represent smaller values of the log mean square error. As we can see, the mean square error is not minimised for a unique pair of flow parameters but along an entire curve. This defines a family of homophonic vortices, i.e. vortex flows with the same characteristic spectrum. Panel D represent the log mean square error computed using the both sides of the light ring spectrum. In this case, the mean square error is minimised at one point in the $(C,D)$ space which is the intersection point of the two homophonic curves obtained from panel C and D. The intersection is located at $(C,D) =(1,1)$ which are the values used to produce the initial spectrum.}
\label{ABHS_deepwater}
\end{figure*}

\textit{Step 3 - Parameter extraction:} 
Once the experimental spectrum is obtained, one can estimate the flow parameters that best match the theoretical model with the experiment by minimizing the mean square error (MSE):
\begin{equation}
MSE(C,D) = \frac{1}{M}\sum_{m} \left(\omega_{\rm exp}(m) - \omega_\star(C,D,m) \right)^2,
\end{equation}
where $M$ is the number of azimuthal components considered.
We consider the mean square error to be a function of the irrotational flow parameters only and we assume that any other parameters entering in the characteristic spectrum (e.g.~the water depth) can be measured independently. 
We stress here that in order to single out a unique pair of flow parameters, one needs to be able to measure both the co ($m>0$) and counter ($m<0$) rotating LR modes. 
Indeed, as can be seen from panels B and C of Figure \ref{ABHS_deepwater}, there is an entire family of flow parameters lying on a curve that minimises the MSE if one only has access to the negative (positive) $m$ part of the spectrum. Since each point in parameter space corresponds to a different vortex flow, these curves determine a family of \emph{homophonic} vortices, i.e. vortex flows with the same characteristic frequency.
It is the intersection of these two homophonic curves that uniquely defines a pair of flow parameters, as shown in panel D of Figure \ref{ABHS_deepwater}.
In the special case of the flow being governed by a single parameter then the negative $m$ part of the spectrum is sufficient to fully characterise the flow. Indeed, in the case of a purely rotational flow (i.e. D=0) there are no co-rotating light-ring modes, while in the case of a purely draining flow (i.e. C=0), the co and counter rotating modes coincide.
\newline

We note that while the general idea of ABHS and BHS is the same, their practical application is very different. 
Indeed, the BHS procedure relies on the matching of a known waveform to the entire signal. 
Generally the waveform will be dominated by one azimuthal number (the least damped one) but the scheme doesn't make explicit use of the symmetry and the information contained in each azimuthal component. 
This is simply due to the fact that in astronomy, one does not have access to the different azimuthal component experimentally. 
This would require many detectors located around the source of the signal, which is not yet realisable. 
The ABHS method makes use of the possibility to measure the perturbation at various positions. 
This allows for the decomposition into the various azimuthal components. 

Another difference is that the ABHS scheme focuses on the real part of the spectrum. 
This is due to the fact that the oscillation frequency is a more robust quantity to look at than the decay rate. 
Indeed, the decay rate depends highly on the boundary conditions of the system. 
Formally, the QNM spectrum is obtained by considering perturbations which are purely outgoing at infinity and ingoing at the horizon. 
These assumptions cannot be satisfied in an experiment; the boundary conditions will modify the imaginary part of the characteristic spectrum (because of the presence of reflections or echoes). 
The real part of the spectrum will however stay unchanged, independently of the boundary conditions.
We note however that if one has access to it, the imaginary part of the characteristic spectrum can be used in conjunction with the real part to single out unique flow parameters. This is illustrated in section \ref{Slow_drain}.
We further note that black holes are not the only objects exhibiting LRs. Because of the local nature of the LRs, the ABHS method is applicable as long as the light rings are located in the irrotational region of the flow, even when the system does not exhibit an analogue horizon.
\\

We now turn to an application of the method: shallow water waves propagating on top of an irrotational DBT flow. 
\section{Application of Analogue Black Hole Spectroscopy} \label{Application}

In this section, we apply our method to a numerically simulated perturbed vortex flow described by the velocity field \eqref{DBT}. 
The wave equation to be solved is equation \eqref{wave_equation} with the $\tanh$ function linearised to first order (shallow water approximation):
\begin{equation} \label{shallowWE}
\matD^2 \phi - c^2 \nabla^2 \phi = 0.
\end{equation}
Exploiting the angular symmetry, this equation can be reduced to a PDE in $(r,t)$ by inserting into the wave equation the ansatz:
\begin{equation}
\phi(r,\theta,t) = \phi_m(r,t) e^{im\theta}.
\end{equation} 
The resulting equation is solved using the Method of Lines (MOL).
To implement this numerically, we define $\pi=\mathcal{D}_t\phi$ to convert wave equation~\eref{shallowWE} into a vector equation which is first order in time,
\begin{equation} \label{discreteWE}
\partial_t \begin{pmatrix}
\phi_m \\ \pi_m
\end{pmatrix} = \begin{bmatrix}
\frac{D}{r}\partial_r-\frac{imC}{r} & 1 \\
c^2(\partial_r^2+\frac{1}{r}\partial_r-\frac{m^2}{r^2}) & \frac{D}{r}\partial_r-\frac{imC}{r}
\end{bmatrix} \begin{pmatrix}
\phi_m \\ \pi_m
\end{pmatrix}
\end{equation}
We then introduce discrete radial points $r_i$ where $i$ is an integer ranging from $1$ to $N$, and approximate the derivatives using 3-point finite difference stencils,
\begin{equation} \label{stencils}
\p_rf_i = \frac{f_{i+1}-f_{i-1}}{2\Delta r}, \qquad \p_r^2f_i = \frac{f_{i+1}-2f_i+f_{i-1}}{\Delta r^2}
\end{equation}
where $f$ is a place holder for $\phi_m$ or $\pi_m$. The boundary condition at $r_N$ is a hard wall, implemented by setting $f_{N+1}=0$, with $r_N$ placed sufficiently far away that reflections do not occur. 
We place a free boundary inside the horizon which is implemented using a one-sided stencil at $r_1$ (the justification for this is that the value of the field inside the horizon cannot effect its value outside since the two regions are causally disconnected). 

We initialise the simulation with a gaussian pulse centred at $r=r_N-5\sigma$, where $\sigma$ is the spread of the gaussian pulse,
\begin{equation} \label{gaussian}
\phi_m(r,t=0) =  \frac{1}{\sqrt{2\pi\sigma^2}}\exp\left(-\frac{[r-r_0]^2}{2\sigma^2}\right),
\end{equation}
with $\pi_m$ chosen such that the perturbation propagates toward $r<0$. 

To extract the real part of the QNM frequencies, we first compute the time Fourier transform $\hat{\phi}_m(\omega)$ of  $\phi_m(t,r=5r_H)$. The location of the peaks in $\hat{\phi}_m$ determines the real part of the QNM spectrum: the peak at $\omega>0$ gives $\omega_\star(m)$ and the peak at $\omega<0$ gives $\omega_\star(-m)$. Hence we only need to simulate for $m>0$. 
At large $m$, the co-rotating modes are much harder to excite, hence the peak at $\omega>0$ becomes difficult to resolve beneath the tail of the $\omega<0$ peak. This accounts for the poor behaviour of the blue crosses at large $m$ on the right of Fig.2. 
Finally, we extract the imaginary part by finding the gradient of $\log|\phi_m(t,r=5r_H)|$ in the region where the signal is exponentially decaying. To do this, we must have one dominant frequency in the signal (i.e. one peak in $\hat{\phi}_m$) since the interference between two frequencies leads to a non-constant gradient.

We focus on two cases: a slowly draining flow and a flow where the circulation and drain parameters are equal. 
After the QNM spectrum is extracted from each simulation, a non-linear regression analysis is performed in order to extract the flow parameters that best match the characteristic spectrum according to the Analogue Black Hole Spectroscopy method. The obtained parameters are compared to the ones used to simulate the data. In addition, we compare our results to a standard flow measurement technique: \emph{Particle Imaging Velocimetry} (PIV).

\subsection{Slowly Rotating Flow}
\begin{figure}[!h]
\begin{center}
\includegraphics[scale=1]{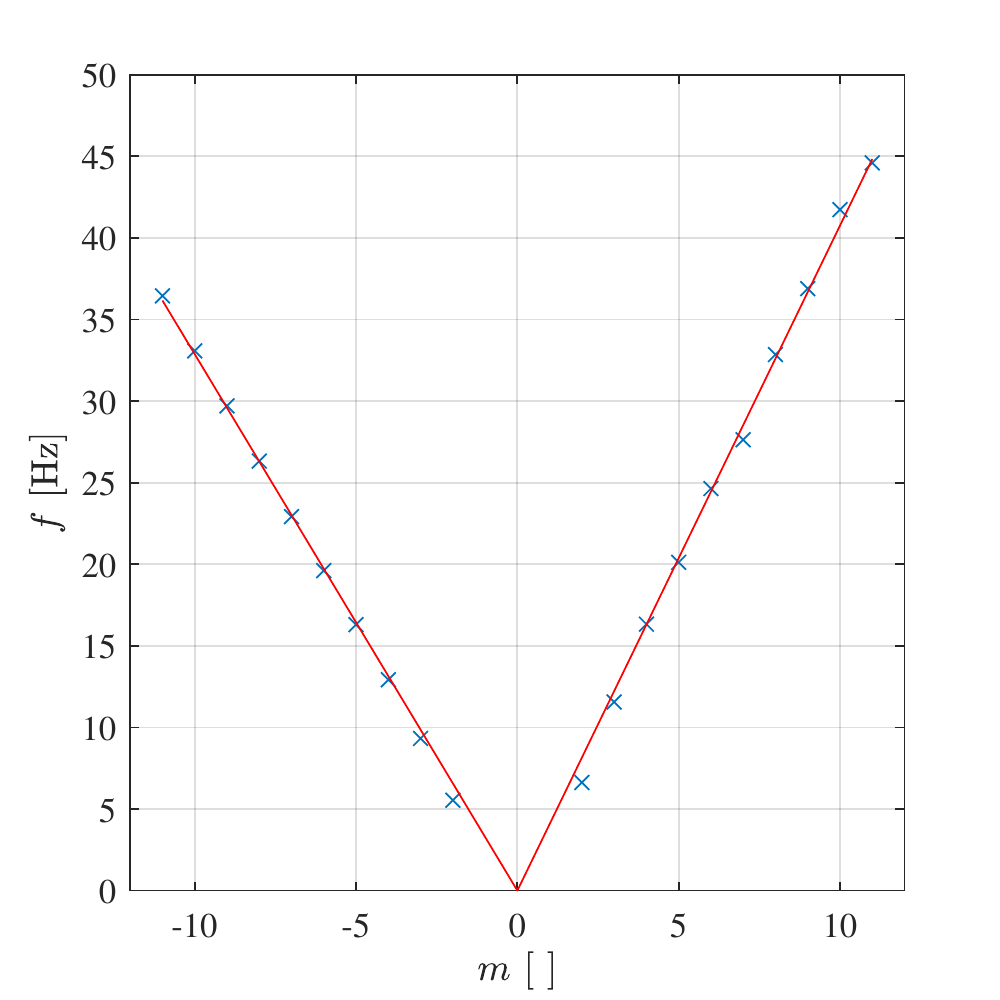}
\end{center}
\caption{ \textbf{Characteristic spectrum of DBT vortex flow in the shallow water regime.} The blue crosses represent the oscillation frequency of the characteristic modes emitted by a draining bathtub flow with \mbox{$C= 2.2\times 10^{-3}~m^2.s^{-1}$} and \mbox{$D=2.2\times 10^{-2}~m^2.s^{-}$} perturbed by a Gaussian wave packet. The red line is the spectrum obtained via the light-ring method with the flow parameters \mbox{$C_\star= 2.3\times 10^{-3}~m^2.s^{-1}$} and \mbox{$D_\star= 2.17\times 10^{-3}~m^2.s^{-1}$}.}\label{shallow_spectrum}

\end{figure}

Here the flow parameters are $h = 10~cm$, \mbox{$C = 2.2\times 10^{-3}~m^2.s^{-1}$}, and \mbox{$D = 2.2\times 10^{-2}~m^2.s^{-1}$}. 
This results in a slowly rotating flow. In this regime, we can see from \eref{shallow_LR} that the co and counter-rotating oscillatory spectrum are located in the same frequency range. Therefore, if one is able to observe one side of the real spectrum, the other side should also be observable.  With these parameters, we are able to extract the characteristic spectrum for $|m| < 11 $, see Figure \ref{shallow_spectrum}.

\begin{figure}[!t]
\begin{center}
\includegraphics[scale=0.8,trim= 2cm 0 2cm 0]{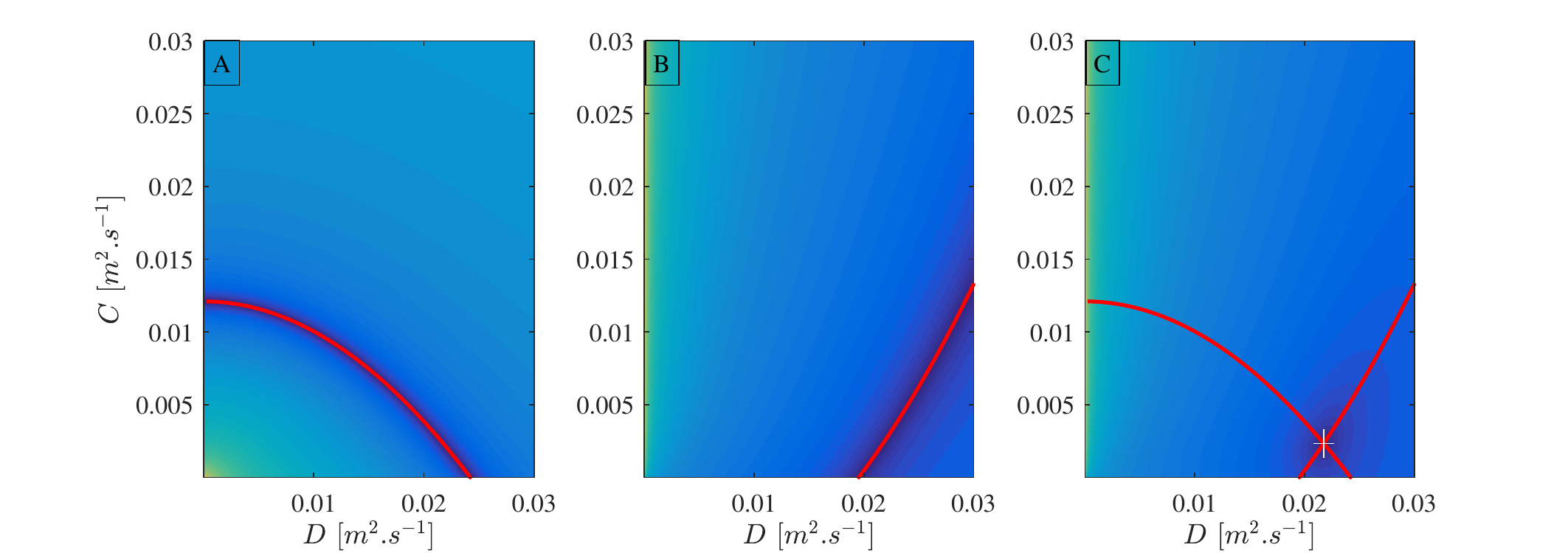}
\end{center}
\caption{ \textbf{Application of the Analogue Black Hole Spectroscopy to numerical simulation: Slowly Rotating Flow}: Panels A, B and C represent the logarithm of the mean square error using the spectrum of negative, positive and all azimuthal numbers $m$ respectively. Darker regions represent smaller values of the log mean square error. Similarly to Panel B and C of figure \ref{ABHS_deepwater}, there is an entire curve in parameter space corresponding to homophonic vortices that minimises the mean square error. These two homophonic curves intersect, as seen in panel C, to give the values of the flow parameters that best match the spectrum. The right cross on panel C depicts the $95\%$ confidence interval and is obtained using the covariance matrix.} \label{NLR_shallow}
\end{figure}

We then apply a non-linear regression analysis to match the numerical spectrum with the formula for the characteristic modes \eref{shallow_LR}, using the positive part of the spectrum, the negative part and the entire spectrum. The mean square error for each analysis is respectively plotted in panels A, B and C of Figure \ref{NLR_shallow}.

Minimising the mean squared error, we find $C_\star = 2.3 \times 10^{-3} \pm 1.0\times 10^{-3}~m^2.s^{-1}$ and $D_\star = 2.17\times 10^{-2} \pm 0.1\times 10^{-2}~m^2.s^{-1}$. The error on the parameters represents the $95\%$ confidence interval and are computed using the covariance matrix. We see that the obtained values are very close to the ones used in the simulation (relative error of about $6\%$ on the estimation of the circulation parameter $C$ and about $1\%$ for the drain parameter). The discrepancy can be explained by the fact that the light-ring method used to estimate the characteristic frequencies is only an approximation (the higher the azimuthal number is, the better the approximation will be). 

Finally, we have extracted the flow parameters using a standard flow visualisation technique called \emph{Particle Imaging Velocimetry} (PIV). To create test data for the PIV method, we seed an initial image with particles of 4 pixel diameter such that the density is $0.013$ particles/pixel (these values were chosen to keep data as close as possible to that obtained in the experiments of \cite{Torres18}) We then evolve the position of the particles using the velocity field in equation~\eref{DBT} to create a second image. These images are analysed using Matlab's extension PIVlab, developed in \cite{thielicke2014pivlab,thielicke2014pivlab2,thielicke2014flapping}. We used PIVlab's window deformation tool with spline deformation to reduce the number of erroneous vector identifications. The analysis was performed over three iterations using an interrogation area of $256\times256$, $128\times128$ and $64\times64$ pixels respectively, each with 50\% overlapping. To obtain $C$ and $D$, we perform a $\theta$ average on the vector field found in PIVlab, then use MATLAB's curve fitting tool to fit the data with a function of the form \eref{DBT}. The flow parameters extracted via this method are: \mbox{$C_{\rm PIV} = 2.2\times 10^{-3} \pm 0.5\times 10^{-4}~m^2/s$} and \mbox{$D_{\rm PIV} = 2.31\times 10^{-2} \pm 0.1\times 10^{-3}~m^2/s$}. Again, the errors on the parameters represent the $95\%$ confidence interval obtained via the covariance matrix. This represents a relative error on the drain parameter of about $2\%$ and on the drain parameter of about $5\%$. We note that while the confidence interval are significantly smaller for the PIV method, a systematic error is present in the method. We further remark than the relative error on the parameters obtained using both methods are of the same order, proving the validity of ABHS as a flow measurement technique.

\subsection{Rotating and Draining Flow}\label{Slow_drain}

\begin{figure}[!h]
\begin{center}
\includegraphics[trim = 2cm 0 0 0]{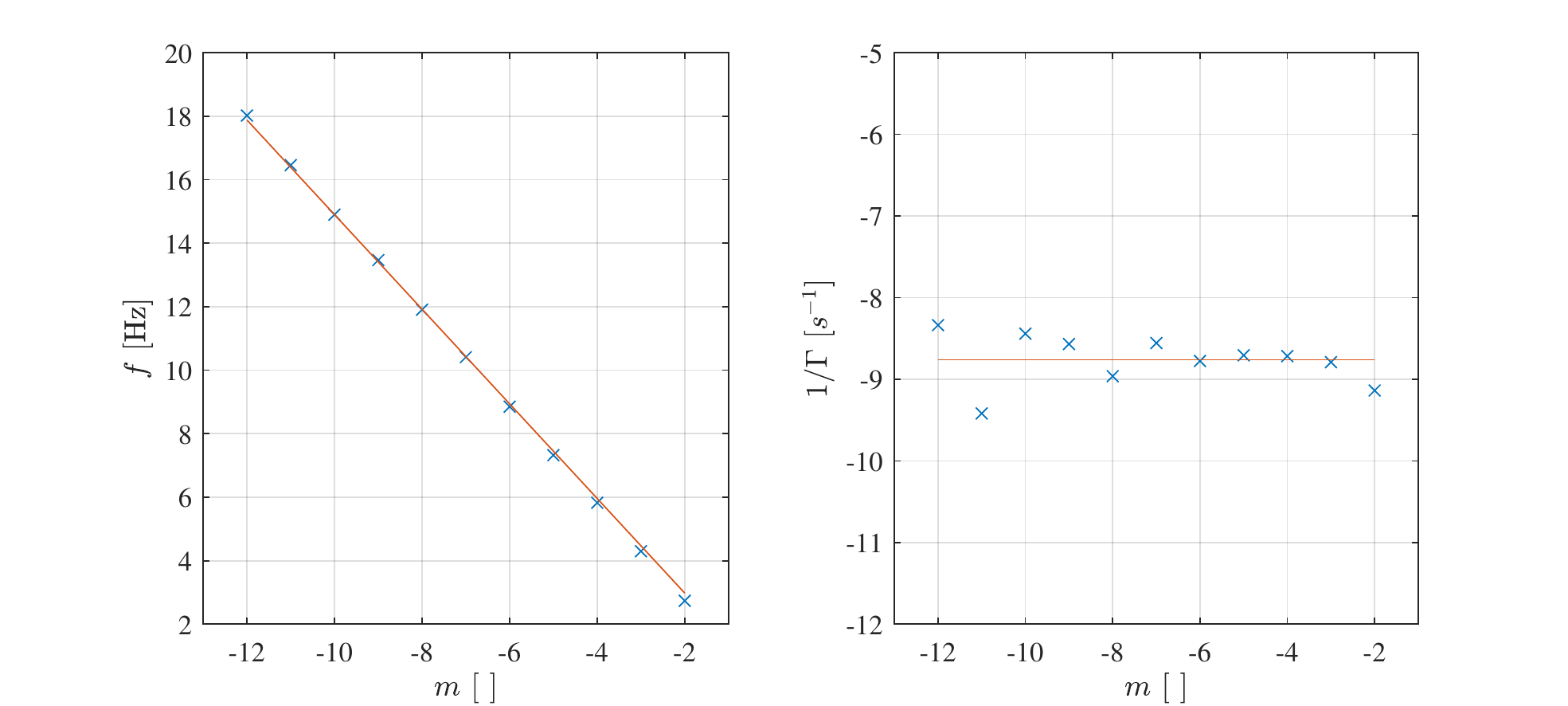}
\end{center}
\caption{ \textbf{Quasinormal mode spectrum of a draining bathtub vortex flow in the shallow water regime}. The left panel represents the oscillation frequencies of the characteristic modes of the draining vortex flow as a function of the azimuthal numbers. The right panel depicts the imaginary part of the characteristic spectrum. In both panels, the blue crosses are the frequencies obtained numerically and the red line represents the best match between the numerical spectrum and the light-ring prediction. The values of the flow parameters used in the simulation are \mbox{$C=D 2.2\times 10^{-2}~m^2.s^{-1}$}. The values that best match the numerical spectrum are \mbox{$C_\star= 2.29\times 10^{-2}~m^2.s^{-1}$} and \mbox{$D_\star=2.01\times 10^{-2}~m^2.s^{-}$}}\label{real_and_im_spec}
\end{figure}

We now turn our attention to a flow where the drain and circulation parameters are of the same order. In this regime, the LR radius of the co-rotating modes will be very close the center of the vortex. 
This will result in a very high oscillatory frequency and short decay time, making the observation of the co-rotating characteristic significantly more difficult. 
However one can use the decay time of the counter-rotating modes in order to single out a unique pair of flow parameters. 
This is the path we adopt here in order to test our method. 
However we note once more that in a real experiment, the imaginary part of the characteristic modes will be highly influenced by the boundary conditions and one should be careful when using the decay time to extract information about the set up.
The flow parameters are here set to be $C = D = 2.2\times 10^{-2}~m^2.s^{-1}$. With these parameters, we can extract the real and imaginary part of the QNM spectrum for $-12<m<-2$. The spectrum is shown in figure \ref{real_and_im_spec}.

Using the real and imaginary part of the spectrum to perform the non-linear regression analysis we find the parameters ($C_\star$,$D_\star$) that best match the spectrum as the intersection of the two curves minimizing the mean square error of each spectra. We find that \mbox{$C_\star = 2.29\times 10^{-2}\pm 0.26\times 10^{-2} ~ m^2/s$}
 and \mbox{$D_\star = 2.01\times 10^{-2}\pm 0.44\times 10^{-2} ~ m^2/s$}, which are relatively close to the values used in the simulation. The relative error on the circulation parameter is about $4\%$ and on the drain parameter is about $9\%$. We note that in this case the resolution on the parameters is worst than in the previous case. This is due to the fact that it is harder to resolve decay times than frequencies. 
 
 Similarly, we compare the ABHS approach with the standard PIV technique. Using PIV the flow parameters are found to be:
 \mbox{$C_{\rm PIV} = 2.29\times 10^{-2} \pm 0.2\times 10^{-3}~m^2/s$} and \mbox{$D_{\rm PIV} = 2.33\times 10^{-2} \pm 0.1\times 10^{-3}~m^2/s$}. This represent a relative error of about $4\%$ for the circulation parameter and about $6\%$ for the drain parameter. As in the previous case, the confidence intervals are significantly smaller when using PIV but a systematic error is present in the result. The relative errors are again comparable between the two techniques.

\section{Discussion} \label{sec:disc}

In this paper we have pursued the idea initiated in~\cite{Torres18} that vortex flows can be identified via their characteristic frequencies.
This method, that we call Analogue Black Hole Spectroscopy, is based on the fluid-gravity analogy and the fact that the quasi-normal mode spectrum of black holes is fully determined by the black hole parameters. 
We have outlined the main steps of the method and applied it to some numerical data to show its potential in real life applications. 
We have also compared our new method with an existing flow measurement technique routinely used in hydrodynamic experiments and called Particle Imaging Velocimetry (PIV). 
We have seen that both methods give similar relative errors when estimating the flow parameters. 
However, the resolution on this estimation is still better when using PIV. 
We believe however that it is possible to increase the resolution of the ABHS method by measuring the characteristic frequencies for a broader range of azimuthal modes. 
Another possibility would be to develop an alternative method applicable to dispersive systems to compute more accurately the characteristic spectrum, similar to the continuous fraction method~\cite{leaver} or via numerical simulation~\cite{pricepullin,krivan}. 
This method, applicable to fluids and superfluids alike, therefore provides a new and non-invasive technique to identify vortex flows. 
It can be used in conjunction with another non-invasive technique based on wave scattering~\cite{Steinberg97,Coutant15} and the fluid analogue of the Aharonov-Bohm effect. 
One can send waves onto a vortex flow and measure the wavefront dislocation to determine the circulation of the flow. 
Then one would observe the vortex response to the perturbation and measure the characteristic spectrum to apply the ABHS scheme.

Furthermore, we have seen that in order to perform the non-linear regression analysis, one needs to be able to compute theoretically the characteristic spectrum. 
In other words, this means that one needs to know the effective field theory applicable to the system. 
This is what we have assumed in this study, and we have shown that from the effective field theory one can reconstruct the flow parameters. 
However, the argument can be applied both ways. If one knows the flow parameters (using other flow measurement techniques) one can constrain the effective field theory describing the system.

Finally, in this paper, we have focussed our attention on an irrotational vortex flow. 
While this system is very interesting theoretically and constitutes a good first model to describe experiments, it remains too naive to capture real hydrodynamical vortices. 
Indeed, due to the divergence of the flow at the centre of the irrotational vortex, one expects the irrotational assumption to break down around some radius. 
Below this radius, the flow becomes rotational and more parameters are needed in order to describe it.
A natural question to ask is therefore, what happens to the ABHS method in the case of more complicated vortex flows?
A first remark to address this question is that LRs are local objects. 
Hence, if the flow is sufficiently fast for the LRs to be in the irrotational regions, the real part of the characteristic spectrum linked with them will not be significantly affected. 
We stress here the fact that the real part will not be considerably affected. 
The imaginary part on the other hand will be modified as the inner boundary condition will be modified by the presence of extra-structures. 
A similar behaviour, manifested by the appearance of echoes, has been predicted in black hole physics when one adds structure (such as a wall) outside the event horizon~\cite{Cardoso16}.
Secondly, if the flow presents extra structures described by extra parameters relevant in the effective field theory, one would expect the system to exhibit extra features. 
For example, the case of shallow water waves incident on a vortex flow with a rotational core has been studied recently~\cite{Patrick18}.
This study, using a simplified model, predicted that the characteristic response to perturbations of this more realistic vortex flow will be composed of the usual ringdown modes together with extra modes, known as quasi-bound states. 
By measuring the quasi-bound state frequencies in addition to the quasi-normal frequencies, one would have access to extra conditions in order to identify the extra parameters needed to describe the flow.

Answering precisely all these questions raised in the paragraph above is beyond the scope of the present paper. Our aim here was to show the potential applicability of the fluid-gravity analogy and lay down the foundations of the ABHS method for future investigations (both theoretically and experimentally) in classical and in quantum fluids.

\ack
The authors thank Sebastian Erne for discussing various aspects of the project.
M.~R.~acknowledges financial support from the S\~ao Paulo Research Foundation (FAPESP, Brazil), Grants No.
2013/09357-9 and 2018/10597-8, and from Conselho Nacional de Desenvolvimento Cient\`{i}fico e Tecnol\'{o}gico (CNPq, Brazil), Grant FA 309749/2017-4. M.~R.~is also grateful to the University of Nottingham for hospitality while this work was being completed. SW acknowledges financial
support provided under the Paper Enhancement Grant at
the University of Nottingham, the Royal Society University
Research Fellow (UF120112), the Nottingham Advanced
Research Fellow (A2RHS2), the Royal Society
Project (RG130377) grants, the Royal Society Enhancement
Grant (RGF/EA/180286) and the EPSRC Project
Grant (EP/P00637X/1). SW acknowledges partial support
from STFC consolidated grant No. ST/P000703/.

\section*{References}
\bibliographystyle{iopart-num}
\bibliography{ABHS_bibli}
\end{document}